\documentclass[article,amsmath,11pt,amssymb,floatfix,superscriptaddress,
showpagenumber,showpacs,nofootinbib]{revtex4}
\usepackage{graphicx}
\usepackage{epsfig}
\usepackage{bm}
\usepackage{amsfonts}
\usepackage{xcolor}
\usepackage{caption,subcaption}
\input epsf

\begin{document}
\title{Cosmology of the interacting Tsallis holographic dark energy in $f(R,T)$ gravity framework}

\author{Sanjeeda Sultana}
\email{sanjeeda.sultana0401@gmail.com; sanjeeda.sultana1@s.amity.edu}
\affiliation{ Department of Mathematics, Amity University, Kolkata, Major
Arterial Road, Action Area II, Rajarhat, Newtown, Kolkata 700135,
India.}

\author{Chayan Ranjit}
\email{chayanranjit@gmail.com}
\affiliation{ Department of Mathematics, Egra S. S. B. College, Purba Medinipur 721429, West Bengal, India.}

\author{Surajit Chattopadhyay}
\email{schattopadhyay1@kol.amity.edu; surajitchatto@outlook.com}
\affiliation{ Department of Mathematics, Amity University, Kolkata, Major Arterial Road, Action Area II, Rajarhat, Newtown, Kolkata 700135,
India.}

\author{Ertan Güdekli}
\email{gudekli@istanbul.edu.tr}
\affiliation{Department of Physics, Istanbul University, Istanbul 34134, Turkey.}
\date{\today}

\newpage
\begin{abstract}
 \begin{center}
 \textbf{Abstract}:
 \end{center}
In this work, we have analyzed the cosmology of the Tsallis holographic dark energy (THDE), a particular case of Nojiri-Odintsov HDE proposed in [S. Nojiri and S. D. Odintsov, \textit{Gen. Relativ. Gravit.} \textbf{38} (2006), 1285; \textit{Eur. Phys. J. C} \textbf{77} (2017) 528], using Hubble's horizon cutoff in $f(R,T)=\mu R+\nu T$ model considering pressureless dark matter. We have examined the equation of state (EoS) parameters in this scenario. The deceleration parameter has been evaluated for this interacting model to justify the late-time acceleration of the expanding universe. We have also studied the cosmological consequences of Statefinder pair, $O_{m}(z)$ diagnostics, $r-q$ plane, and $w_{DE}-w^{'}_{DE}$ pair for interacting THDE in $f(R,T)=\mu R+\nu T$ model. We have also illustrated the cosmology of the interacting THDE  using Hubble's horizon cutoff in $f(R,T)=R+\gamma R^2+\xi T$ model. The EoS parameter, deceleration parameter and Statefinder pair are studied in this interacting scenario. Attainment of $\Lambda$CDM fixed point has been observed for both models. We have also constrained model parameters based on observational data sets through the formalism of $\chi^{2}$ minimum test.
\end{abstract}

\maketitle

\section{Introduction}
One of the most important issues in contemporary cosmology is the origin of dark energy (DE) \cite{C1,C2,C3}, which is responsible for the cosmic acceleration \cite{C4,C5}. Determining whether DE is a straightforward cosmological constant or if it comes from various sources that dynamically change over time \cite{C6} is the first step towards comprehending its nature. Nojiri and Odintsov \cite{C7} provided a comprehensive overview of the principles that make the modified gravity technique so appealing for applications involving DE and the late time acceleration of the universe. Here are several other outstanding reviews on modified gravity \cite{C6,C8}. Several modified gravity theories have been put forth so far \cite{52,53,54,55,56,57}. These consist of the following theories: Horava-Lifshitz \cite{C9,C10}, Gauss-Bonnet \cite{C11,C12}, $f(R)$ \cite{C13,C14}, $f(T)$ \cite{C15,C16,C17,C18}, and $f(G)$ \cite{C19,C20}. The current work focuses on $f(R,T)$ gravity, where $T$ is the stress-energy tensor trace, indicating a coupling between geometry and matter. Let's quickly look at the $f(R)$ gravity before delving into the details of $f(R,T)$ gravity. The need to explain the Universe's apparent late time accelerating expansion \cite{C8} has recently spurred interest in $f(R)$ gravity research. A few in-depth analyses of $f(R)$ gravity are found in \cite{C21,C22,C23,C24}. In the works of \cite{C25,C26}, thermodynamic aspects of $f(R)$ gravity have been studied. 

A generalization of $f(R)$ modified
theories of gravity includes an absolute coupling of an arbitrary function of the Ricci
scalar $R$ with the matter Lagrangian density $\mathcal{L}_{m}$ that arouses motion of the massive particles which is non-geodesic, and an extra force which is orthogonal to the four-velocity \cite{C27,C28}. The authors of \ proposed an extension of standard general relativity cite{C27}, where the gravitational Lagrangian is given
by an arbitrary function of the Ricci scalar $R$ and of the trace of the stress-energy tensor $T$ and named it as $f(R, T)$ gravity. A source term that reflects the variation of the matter stress-energy tensor with respect to the metric is necessary for the $f(R,T)$ gravity model to function. This source term's general expression is derived as a function of the matter Lagrangian $\mathcal{L}_{m}$ \cite{27}. Exact solutions for a specific model $f(R,T)=\mu R+\nu T$ \cite{C29,C029} have been derived and showed that we can have accelerated expansion of the universe for some values of the parameters without initiating any dark component.  

Two theories extensively studied in the literature to explain the late time accelerated expansion are modified gravity and DE theory. Reviews of the DE and modified gravity theories to account for the cosmic acceleration in late time can be found in \cite{C30, C31, C32, C33, C34, C35}. A viable method for resolving the DE conundrum that has drawn a lot of interest is the holographic DE (HDE) hypothesis \cite{C36,C036,C37,C38,C39,C40,C41,C42,C43,C043,C44}. Furthermore, a flat Friedmann-Robertson-Walker (FRW) \cite{C036,C37,C39} universe's history cannot be well described by the fundamental model of HDE, which is based on the Bekenstein entropy and the Hubble horizon as its IR cutoff. In an attempt to overcome these shortcomings, physicists take into account: (i) alternative cutoffs; (ii) potential interactions between the cosmic sectors; (iii) different entropies; or perhaps a mix of the aforementioned strategies \cite{C043,C44}. Many generalized entropy formalisms have been used to analyze cosmological and gravitational phenomena \cite{C45,C46,C47,C48,C49,C50,C51,C52,C53,C54} because of the long-range nature of gravity and the uncertainty surrounding spacetime.  These investigations' findings suggest that the power-law distributions of probability (generalized entropy formalisms) agree reasonably with gravity and its related issues.

Two additional HDE models have been proposed recently \cite{C55,C56}, assigning different generalized entropies to the FRW universe's horizon. The foundation of these efforts is the realization that the Bekenstein entropy may be acquired by applying Tsallis statistics to the system horizon\cite{C57,C58,C59,C60}. Additionally, the obtained models independently exhibit acceptable stability\cite{C55,C56}. The Tsallis definition of entropy \cite{C61} is extremely important for understanding the gravitational and cosmological systems within the context of generalized statistical mechanics\cite{C45,C46,C47,C48,C49,C50,C51,C52,C53,C54,C55,C56,C61,C62}. Tsallis and Cirto \cite{C61} have demonstrated that the system's Tsallis statistics yield more results than the Bekenstein entropy. Quantum gravity considerations \cite{C63} reveal that the Tsallis entropy content of a system is, in general, a power-law function of the system's area \cite{C61}. It should be noted that  Tsallis HDE is just a particular example of Nojiri-Odintsov HDE proposed in \cite{Nojiri:2005pu, Nojiri:2017opc}. It was explicitly stated in \cite{Nojiri:2021iko1,Nojiri:2021iko2} that Tsallis HDE is a particular example of NO HDE. A more generalized Tsallis HDE was reported in \cite{Nojiri:2019skr}. Still, above HDE is primer of Nojiri-Odintsov HDE.

Numerous dynamic DE theories are being developed to explain the universe's accelerated expansion, so a delicate test is necessary to distinguish between these DE models. Sahni et el. \cite{8} and Alam et el. \cite{9} thus put out a scientific recommendation that affects the application of the parameter to coordinate $(r,s)$, the specified ``Statefinder." With higher-order differential coefficients of the scale factor $a$, it is easy to find the Statefinder parameter to illuminate the components of the universe's evolution. This becomes a trademark to pursue stage past the deceleration parameter $q$ and Hubble parameter $H$. Given that different cosmological models with DE exhibit theoretically distinct advancement directions in the $s-r$ plane, Statefinder parameter is a useful approach to observe these DE models. Since model-dependent physical variables describe dark energy characteristics, Statefinder analysis is formulated directly from a space-time metric, making it more universal. Generally, the evolutionary trajectories are shown in the $r-s$ plane to visualize the subjectively distinct practices corresponding to the various dark energy models.  A fixed point $(r=1,s=0)$ corresponds to the spatially flat $\Lambda$CDM scenario. One convenient method to measure the displacement of a given model from $\Lambda$CDM is to calculate the distance between this fixed point and the given dark energy model. According to \cite{9,I0,I1,I2}, the statefinder can completely differentiate among various DE models, such as the Chaplygin gas models, the cosmological constant, quintessence, braneworld and interacting DE models. For different interacting DE models, \cite{I3,I4,I5} have investigated the Statefinder analysis. The authors of \cite{43} proposed the $w_{DE}-w^{'}_{DE}$ plane where $w^{'}_{DE}$ is the derivative of $w_{DE}$ with respect to $\ln a$, to illustrate the dynamical properties of DE models. \cite{I6} studied the Statefinder and $w_{DE}-w^{'}_{DE}$ analysis using agegraphic DE models in both interaction and non-interaction scenarios. 

In the present work, we aim to reconstruct the Tsallis HDE (THDE)\cite{C61} model with Hubble’s horizon cutoff in $f(R,T)$ Gravity by taking the interacting scenario into account to describe the late time acceleration of the universe. Considering modified gravity models with holographic background fluid is well documented in the literature. Karami et al. \cite{reconst1} reconstructed the holographic $f(T)$-gravity model with the power-law entropy correction using the relationship between the $f(T)$-gravity model and the holographic dark energy model.  They \cite{reconst1} used the most recent observational data, such as information on type Ia supernovae, baryon acoustic oscillations, the cosmic microwave background, and Hubble parameter data, to fit the model parameters. In another study, \cite{reconst2} produced a scenaand deceleration parametersk matter and developed the reconstruction of the $f(T)$ gravity model based on holographic dark energy. The non-viscous and viscous holographic dark energy models under modified gravity, where the Hubble horizon sets the infra-red cutoff, were examined in \cite{reconst3}. For the viscous and non-viscous models, \cite{reconst3} discovered the power-law and exponential forms of the scale factor, respectively. Another very relevant study in this respect is by \cite{reconst4}, where a problem similar to what is addressed in the current study was addressed. In that study, they analysed the detailed cosmology of $f(R,T)$ gravity in Tsallis holographic framework. However, the current work has completely deviated from the work of \cite{reconst4} in the sense that in this study, we have considered interacting scenarios, and linear as well as non-linear $f(R,T)$ models are analyzed thoroughly in Tsallis holographic framework. Moreover, a detailed analysis of the model is presented for observational data. Elaborations would be made in the subsequent sections.
The paper is organized as follows: In Section
II, we have briefly reviewed $f(R,T)$ gravity. In Section III, we have presented an overview of the THDE model. In Section IV, we have reconstructed the equation of state (EoS) parameter and deceleration parameter for interacting THDE with Hubble's horizon cutoff in $f(R,T)=\mu R+\nu T$ model. In Section V, we have analyzed the Statefinder pair and $O_{m}(z)$ diagnostics and also the $r-q$ plane. In Section VI, the analysis of $w_{DE}-w^{'}_{DE}$ pair is done.In Section VII, we have studied the cosmology of interacting THDE with Hubble's horizon cutoff in $f(R,T)=R+\gamma R^2+\xi T$ model. In Section VIII, mathematical foundations for observational constraints are presented, and $\chi^{2}$ minimization procedure is implemented for different observational data sets in Section IX. We have concluded in Section X.

\section{$f(R,T)$ Model}
Among all the models of $f(R,T)$ gravity, a more interesting and general form of $f(R,T)$ gravity is the so-called $M_{37}-model$. The action of the $M_{37}-gravity$ \cite{C29,C029} is given by
\begin{equation}
S_{37}=\int d^{4}x\sqrt{-g}[f(R,T)+\mathcal{L}_{m}]
    \label{A1}
\end{equation}
where $f(R,T)$ is an arbitrary function of the scalar curvature $R$ and the torsion scalar $T$ and $\mathcal{L}_{m}$ is the matter Lagrangian. The standard forms of the curvature and torsion scalars are
\begin{equation}
R=u+6(\dot{H}+2H^{2})
    \label{A2}
\end{equation}
and
\begin{equation}
T=v-6H^{2}
    \label{A3}
\end{equation}
respectively, where  $u$ and $v$ are taken as $u = \alpha a^{n}$ and $v = \beta a^{m}$ with $m$, $n$, $\alpha$
and $\beta$ as real constants. The following model of $f(R, T)$ is considered in \cite{C29,C029}
\begin{equation}
f(R,T)=\mu R+\nu T,
    \label{A03}
\end{equation}
where $\mu$ and $\nu$  are real constants. The equation system of this $f(R,T)$ model is
\begin{equation}
\mu D_{1}+\nu E_{1}+K(\nu T+\mu R)=-2 a^{3} \rho
    \label{A4}
\end{equation}
\begin{equation}
\mu A_{1}+\nu B_{1}+M(\nu T+\mu R)=6 a^{2} p
    \label{A04}
\end{equation}
\begin{equation}
\dot{\rho}+3H(\rho+p)=0.
    \label{A004}
\end{equation}
where we have,
\begin{equation}
D_{1}=a^{3}\biggl(6\frac{\ddot{a}}{a}+\dot{a}u_{\dot\alpha}\biggr)
    \label{A5}
\end{equation}
\begin{equation}
E_{1}=a^3\biggl(-12\frac{\dot{a}^{2}}{a^{2}}+ \dot{a}v_{\dot{\alpha}}\biggr)
    \label{A6}
\end{equation}
\begin{equation}
K=-a^{3}
    \label{A7}
\end{equation}
\begin{equation}
    A_{1}=12\dot{a}^{2}+6a\ddot{a}+3a^{2}\dot{a}u_{\dot{\alpha}}+a^{3}\dot{u}_{\dot{\alpha}}-a^{3}u_{\alpha}
    \label{A8}
\end{equation}
\begin{equation}
B_{1}=-24\dot{a}^{2}-12a\ddot{a}+3a^{2}\dot{a}v_{\dot{\alpha}}+a^{3}\dot{v}_{\dot{\alpha}}-a^{3}v_{\alpha}
    \label{A9}
\end{equation}
\begin{equation}
M=-3a^{2}
    \label{A10}
\end{equation}
\begin{equation}
R=u+6(\dot{H}+2H^{2})
    \label{A11}
\end{equation}
\begin{equation}
T=v-6H^{2}.
    \label{A12}
\end{equation}
Hence, we get the modified field equations as \cite{C29,C029}
\begin{equation}
3(\mu+\nu)H^{2}+\frac{1}{2}(\mu\alpha a^{n}+\nu \beta a^{m})=\rho
    \label{A13}
\end{equation}
\begin{equation}
(\mu+\nu)(2\dot{H}+3H^{2})+\frac{\mu \alpha (n+3)}{6}a^{n}+\frac{\nu \beta(m+3)}{6}a^{m}=-p
    \label{A14}
\end{equation}

\section{THDE Model}
The definition and derivation of standard holographic dark energy (HDE) density ($\rho_{D}=3c^{2}m_{p}^{2}L^{-2}$) depends on the entropy–area relationship $S\sim A\sim L^{2}$ of black holes, where $A=4\pi L^{2}$ represents the area of the horizon \cite{C36}. However, due to the quantum considerations \cite{C043,C44}, this definition of HDE can be modified. It was shown by Tsallis and Cirto \cite{C61} that the horizon entropy of a black hole may be modified as
\begin{equation}
S_{\delta}=\gamma A^{\delta},
    \label{A014}
\end{equation}
where $\delta$  and $\gamma$ are non-additivity parameter and an unknown constant respectively \cite{C61}. The Bekenstein entropy is retrieved at the appropriate limit of $\delta=1$  and $\gamma=\frac{1}{4G}$ (in the unit where $h=\kappa_{B}=c=1$). The system may be described by the ordinary probability distribution of probability  and at this limit, the power-law distribution of probability is meaningless. This relationship is further supported by quantum gravity \cite{C63} and yields intriguing outcomes in holographical and cosmic configurations \cite{C51,C54,C64,C65,C66}. Based on the holographic principle, which stipulates that an infrared cutoff should limit a physical system's degrees of freedom and that those degrees should scale with the system's enclosing area rather than its volume \cite{10,11}. A relationship between the system entropy $(S)$ and the UV ($\Lambda$) and IR $(L)$ cutoffs was postulated by Cohen et al. \cite{C36}
\begin{equation}
L^{3}\Lambda^{3}\leq(S)^{\frac{3}{4}}.
    \label{A015}
\end{equation}
On combining Eq.(\ref{A014}) with Eq.(\ref{A015}), we have \cite{C36}
\begin{equation}
\Lambda^{4}\leq\gamma(4\pi)^{\delta}L^{2\delta-4},
    \label{A016}
\end{equation}
where $\Lambda^{4}$ corresponds to the vacuum energy density, the energy density of DE $(\rho_{D})$ in the HDE hypothesis \cite{C43,C67,C68}. The THDE is proposed using the above inequality as
\begin{equation}
\rho_{D}=BL^{2\delta-4},
    \label{A017}
\end{equation}
where $B$ is a parameter \cite{C43,C67,C68}.

\section{Interacting THDE Model with Hubble's Horizon Cutoff in $f(R,T)=\mu R+\nu T$ model }
We have considered THDE interacting with pressureless dark matter in the present work.  Various forms of “interacting” dark energy models have been constructed to fulfill the observational requirements. A Plethora of works of literature is available where the interacting dark energies have been demonstrated. In \cite{1,2,3,4,5,6}, several instances of interacting dark energy are presented. This section considers interacting THDE in $f(R,T)$ gravity. The metric of a spatially flat homogeneous and isotropic universe in the Friedman-Robertson-Walker model is
\begin{equation}
ds^{2}=dt^{2}-a^{2}(t)[dr^{2}+r^{2}(d\theta^{2}+sin^{2}\theta d\phi^{2})],
    \label{E0}
\end{equation}
where $a(t)$ is the scale factor. The Einstein field equations are 
\begin{equation}
H^{2}=\frac{1}{3}\rho
    \label{P1}
\end{equation}
and
\begin{equation}
\dot{H}=-\frac{1}{2}(\rho+p),
    \label{P2}
\end{equation}
where $p$ and $\rho$ are isotropic pressure and energy density, respectively. Here we have chosen $8\pi G = c = 1$. The conservation equation is given by Eq.(\ref{A004}). As the interaction between THDE and dark matter has been considered, we have
\begin{equation}
    \rho=\rho_{DE}+\rho_{m},
    \label{P4}
\end{equation}
\begin{equation}
       p=p_{DE}.
    \label{P04}
\end{equation}
As we are considering pressureless dark matter, so $p_{m}=0$. We must reconstruct the conservation equation by adding an interaction factor $Q$ because the components do not satisfy the conservation equation independently in the case of an interaction. Any one of the following forms \cite{7} could be the interaction term: $Q \propto H(\rho_{DE} +\rho_{m})$,
$Q \propto H  \rho_{DE}$ and $Q \propto H  \rho_{m}$. 
in the present case we have chosen the interaction term $Q \propto H\rho_{m}$ and hence $Q = 3H\delta \rho_{m}$ is our choice where $\delta$ is the coupling constant. The conservation equation is reconstructed as follows:
\begin{equation}
    \dot{\rho}_{DE}+3H(\rho_{DE}+p_{DE})=3H\delta\rho_{m},
    \label{P5}
\end{equation}
\begin{equation}
\dot{\rho}_{m}+3H\rho_{m}=-3H\delta\rho_{m}.
    \label{P6}
\end{equation}
The form of the energy density of Tsallis holographic dark energy (THDE) \cite{C68} considered here is
\begin{equation}
\rho_{DE}=L^{2\xi-4},
    \label{E1}
\end{equation}
where $\xi$ is a constant. The energy density of THDE corresponds to the energy density of HDE model for $\xi=1$.  Let us suppose a flat FRW universe in which the DE candidate does not interact with other portions of the universe, and the Hubble horizon is a suitable candidate for the IR cutoff, i.e., $L=\frac{1}{H}$. So, we get the THDE density as
\begin{equation}
\rho_{DE}=H^{4-2\xi}.
\label{E2}
\end{equation}
From Eq.(\ref{P6}), we get the density of dark matter in the interacting scenario as
\begin{equation}
\rho_{m}=\rho_{m0}a^{-3(1+\delta)}.
    \label{E3}
\end{equation}
Using Eqs. (\ref{E2}) and (\ref{E3}) in Eq.(\ref{A13}) and by taking $\xi=0.5$ and considering $m=n$ for simplicity in calculations, we get the expression for Hubble parameter as [see Eq.(\ref{E4}) of appendix].
Using Eq.(\ref{E4}) in Eq.(\ref{E2}), we get  the energy density for THDE while interacting with
pressureless dark matter in $f(R,T)=\mu R+\nu T$ model as [see Eq.(\ref{E5}) of appendix].

\begin{figure}
\begin{center}
\includegraphics[height=3.5in]{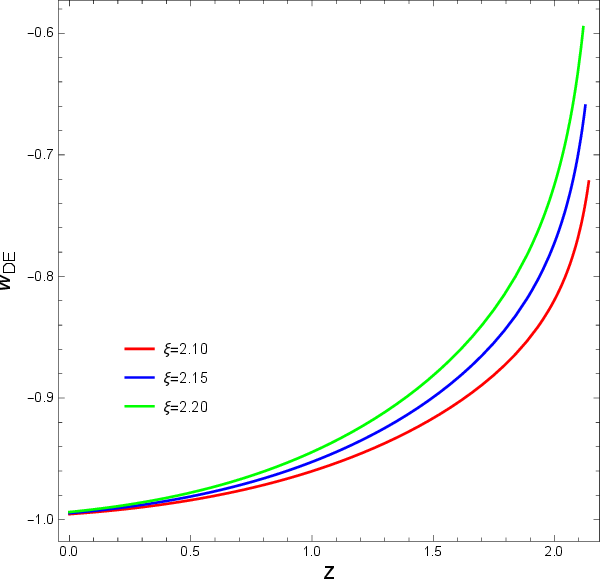}
\caption{Evolution of EoS parameter $w_{DE}$ with respect to redshift $z$ for $f(R,T)=\mu R+\nu T$ model. We have chosen $\rho_{m0} = 0.1$, $\nu = 0.001$, $\mu=1$, $\alpha=0.4$, $\delta=0.03$, $n=0.01$, $\beta=0.3$, $C_{1}=0.5$.}
\label{F1}
\end{center}
\end{figure}

\begin{figure}
\begin{center}
\includegraphics[height=3.5in]{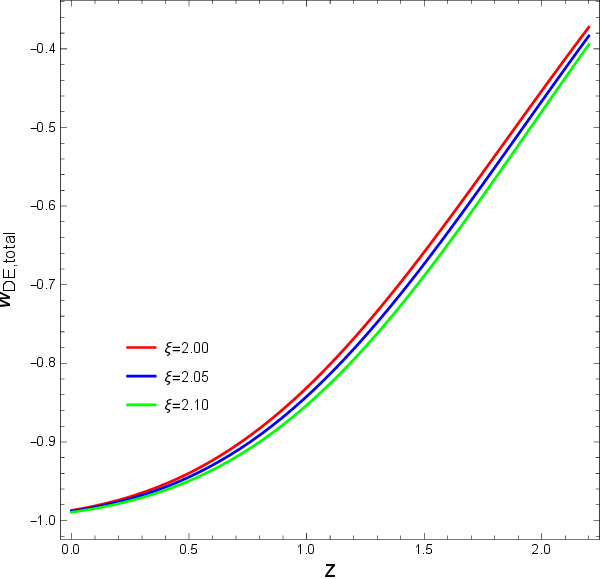}
\caption{Evolution of total EoS parameter $w_{DE,total}$ against redshift $z$ for $f(R,T)=\mu R+\nu T$ model. We have chosen $\rho_{m0} = 0.01$, $\nu = 0.001$, $\mu=0.01$, $\alpha=0.4$, $\delta=0.3$, $n=0.1$, $\beta=0.003$, $C_{1}=0.5$.}
\label{F01}
\end{center}
\end{figure}

\begin{figure}
\begin{center}
\includegraphics[height=3.5in]{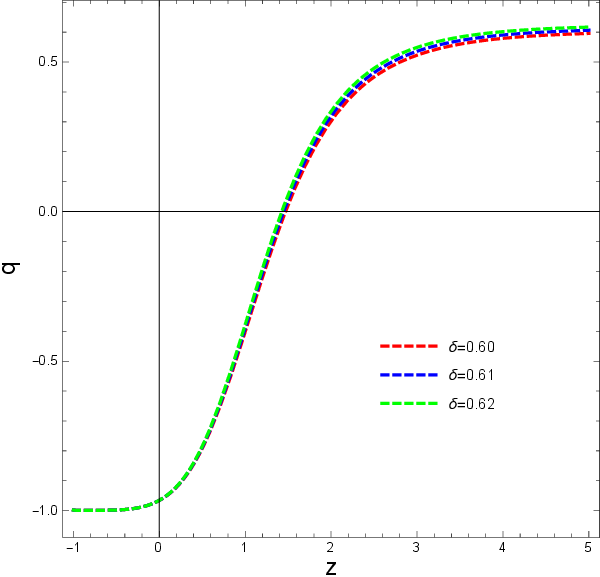}
\caption{Evolution of deceleration parameter $q$ with respect to redshift $z$ for $f(R,T)=\mu R+\nu T$ model. We have chosen $\rho_{m0} = 0.001$, $\nu = 0.01$, $\mu=0.001$, $\alpha=0.004$, $C_{1}=0.05$, $n=0.1$, $\beta=0.3$.}
\label{F2}
\end{center}
\end{figure}

By using Eqs. (\ref{E3}), (\ref{E4}) and (\ref{E5}) in Eq.(\ref{P5}), we have pressure $p_{DE}$ for THDE in the present scenario . From the expressions derived for energy density and pressure of THDE, we have the EoS
parameters 
\begin{equation}
w_{DE}=\frac{p_{DE}}{\rho_{DE}},
    \label{N1}
\end{equation}
\begin{equation}
    w_{DE,total}=\frac{p_{DE}}{\rho_{DE}+\rho_{m}}.
    \label{N2}
\end{equation}
The EoS parameter $w_{DE}$ for interacting THDE with Hubble’s Horizon cutoff in $f(R,T)=\mu R+\nu T$ model is plotted in Fig.\ref{F1} with respect to redshift $z$. From Fig.\ref{F1}, we conclude that it is showing a quintessence behaviour and is decreasing with the evolution of the universe but it is not crossing the phantom boundary $w_{DE} \approx -1$. It also shows that at the early age of the universe, approximately at $z > 2$, the EoS parameter $w_{DE}$ approaches $0$, i.e., in the case of THDE in $f(R,T)$ gravity, the dark energy behaves like dust matter during most of the matter-dominated era. We have also plotted the total EoS parameter $w_{DE,total}$ with respect to redshift $z$ for interacting THDE with Hubble’s horizon Cutoff in $f(R,T)=\mu R+\nu T$ model in Fig.\ref{F01}. From Fig.\ref{F01}, we conclude that it is showing a quintessence kind of behaviour and a decreasing function of cosmic time $t$ in Gyrs but it is not crossing the phantom boundary $w_{DE,total} \approx -1$. 
\begin{table}[ht!]
\caption{The tabular presentation of the value of EoS parameter for the interacting THDE with Hubble’s Horizon Cutoff in $f(R,T)=\mu R+\nu T$ model for the choice of parameters at the point $z=0$ i.e., current accelerating universe.}
\centering
 \begin{tabular}{||c | c |c||} 
 \hline
$(\rho_{m0}, \nu, \mu, \alpha, \delta, n, \beta, C_{1}, \xi)$ & $w_{DE}$ & $w_{Observed} \cite{W0}$\\ [1ex] 
 \hline\hline
 (0.1, 0.001, 1, 0.4, 0.03, 0.01, 0.3, 0.5, 2.1)  & -0.99 & $-1.06_{-0.13}^{+0.11}$ \\[1ex]
 \hline\hline
 \end{tabular}
 \label{T1}
\end{table}
The evolution of the deceleration parameter $q$ against the redshift $z$ for interacting THDE with Hubble's horizon cutoff in $f(R,T)=\mu R+\nu T$ model is given by, 
\begin{equation}
q=-\frac{a\ddot{a}}{\dot{a}^{2}}=-1-\frac{\dot{H}}{H^{2}},
    \label{E6}
\end{equation}
and plotted in Fig.\ref{F2}. From Fig.\ref{F2}, we observe that at the very early phase of the universe, $q > 0$, roughly around $z > 1.5$,  i.e. the decelerated expansion phase of the universe. At $z \approx 1.5$, a transition is seen in the case of  
the deceleration parameter $q$  from a positive to a negative level, which means the universe gradually transits from the decelerated expansion phase $(q > 0)$
to the accelerated expansion phase $(q < 0)$ and at a later stage it converges to
-$1$ and becomes asymptotic in its neighborhood of -$1$. Thus, we conclude that it is possible to have the accelerated expansion phase of the universe from the decelerated expansion phase for interacting THDE with Hubble’s Horizon Cutoff in $f(R,T)=\mu R+\nu T$ model. 

\section{Statefinder and $O_{m}(z)$ Diagnostics}
The cosmological diagnostic pair $(r,s)$ \cite{8,9} depends upon the scale factor $a$ directly and hence upon the metric describing space-time, so in that sense, it is considered “geometrical”. The parameters $r$ and $s$ are defined in terms of Hubble parameter $H(z)$ and its time derivatives as
\begin{equation}
r=1+3\frac{\dot{H}}{H^{2}}+\frac{\ddot{H}}{H^{3}}
    \label{E7}
\end{equation}
and
\begin{equation}
s=-\frac{3H\dot{H}+\ddot{H}}{3H(2\dot{H}+3H^{2})}
    \label{E8}
\end{equation}
respectively. Different DE models \cite{8,9} represent different combinations of $r$ and $s$, such as:
\begin{itemize}
\item For $(r=1,s=0)\rightarrow\Lambda$CDM model
\item For $(r>1,s<0)\rightarrow$ CG model
\item For $(r=1,s=1)\rightarrow$ SCDM model
\item For $(r<1,s>0)\rightarrow$ Quintessence region
\item For $(r=1,s=\frac{2}{3})\rightarrow$ HDE model
\end{itemize}
The holographic dark energy (HDE) model's evolutionary paths in the $s-r$ plane \cite{10,11,12,13,14,15,16} with the future event horizon as the IR cut-off begins at $s=\frac{2}{3},r=1$ and eventually approach the $\Lambda$CDM fixed point $(s = 0, r = 1)$ at a late time \cite{15}. For both the Ricci DE (RDE) model and the quintessence DE model with a constant EoS parameter \cite{8,9}, the $s-r$ plane curves are vertical \cite{17}. In Chaplygin gas (CG), the trajectory in the $s-r$ plane lies in the regions $s < 0, r > 1$ \cite{18}. In contrast, the phantom model with power law potential and the quintessence (inverse power-law) models $(Q)$ lie in the regions $s > 0, r < 1$ \cite{8,9}, and both scenarios approach the $\Lambda$CDM fixed point at late time. In the coupled quintessence models \cite{19}, the trajectory in the $s-r$ plane generates a swirl before arriving at the attractor. At the early phase of the universe, $\Lambda$CDM behaviour is shown in the Agegraphic DE model \cite{20} and the Polytropic gas model \cite{21}. The ghost DE model and the HDE model of DE with model parameter $c = 1$ both exhibit similar kind of behaviour in the $(s, r)$ plane \cite{22}.
This behaviour is likewise consistent with models of DE such as generalized Chaplygin gas \cite{23,24,25}, Chaplygin gas \cite{26,27}, Yang–Mills \cite{28}, new agegraphic \cite{20,29}, and HDE \cite{14,15,16}. The curve of the $s-r$ plane crosses through the $\Lambda$CDM fixed point at the middle of the universe's evolution in the case of the tachyon DE model \cite{30} and HDE model with Granda–Oliveros IR cut-off\cite{31}. In the case of the THDE model, the trajectories of the $s-r$ plane terminate at the $\Lambda$CDM fixed point $(s = 0, r = 1)$ at late time. They begin at the matter-dominated (SCDM) $s = 1, r = 1$ and go through an arc segment and a parabola (downward)\cite{32,33}. The Chaplygin gas behaviour in the case of a RHDE model \cite{34,034} is depicted by the evolutionary curve of the $s-r$ plane, which begins and finishes with a swirl at the $\Lambda$CDM fixed point $(s = 0, r = 1)$. In the late time evolution of the universe, the Statefinder pair $(r,s)$ of the SMHDE model approaches the $\Lambda$CDM fixed point $(r = 1, s = 0)$ and always lies in the Chaplygin gas area, according to recent research by one of the authors \cite{35}. The evolution of the $(r,s)$ pair for the NGCG model has been studied in \cite{36}. In the case of the Tsallis agegraphic dark energy model, the $s-r$ plane's evolutionary curve begins at a cosmological constant, circles a corner, and moves towards a different endpoint \cite{37}. On the other hand, the writers of \cite{38,40} have thoroughly covered the Statefinder pair analysis for a variety of DE models.

\begin{figure}
\begin{center}
\includegraphics[height=3.5in]{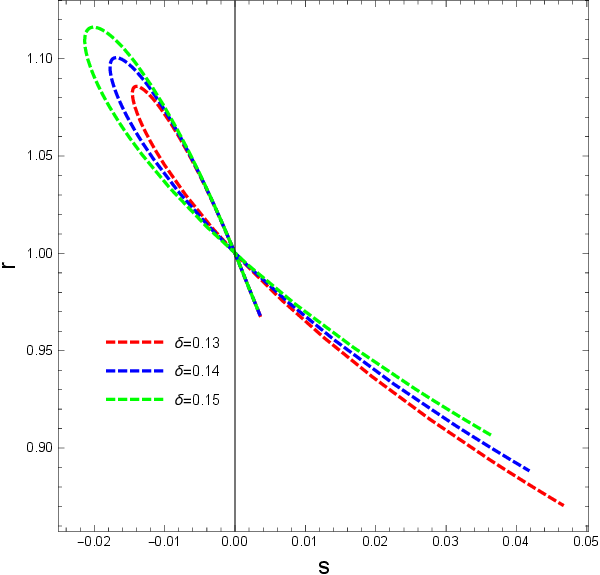}
\caption{The statefinder pair $(r,s)$ trajectory for $f(R,T)=\mu R+\nu T$ model. We have chosen $\rho_{m0} = 0.003$, $\nu = 0.009$, $\mu=0.2$, $\alpha=0.035$, $C_{1}=0.05$, $n=0.1$, $\beta=0.3$.}
\label{F3}
\end{center}
\end{figure}

The evolution of the reconstructed Statefinder pair $(r,s)$ for interacting THDE with Hubble’s Horizon cutoff in $f(R,T)=\mu R+\nu T$ model has been plotted in Fig.\ref{F3}. From Fig.\ref{F3}, we can conclude that the evolutionary trajectories
of the Statefinder pair of the THDE model start its evolution from the region $(s>0,r<1)$ and passes
through the $\Lambda$CDM fixed point $(s=0,r=1)$ as time passby. After making a swirl in the Chaplygin gas region i.e., $s<0,r>1$, it again passes through the $\Lambda$CDM fixed point $(s=0,r=1)$ and remains in the region $(s>0,r<1)$ in future. As the trajectories pass the $\Lambda$CDM fixed point twice, it confirms strongly that the interacting THDE in $f(R,T)=\mu R+\nu T$ model circulates the $\Lambda$CDM phase of the universe.

\begin{figure}
\begin{center}  
\includegraphics[height=3.5in]{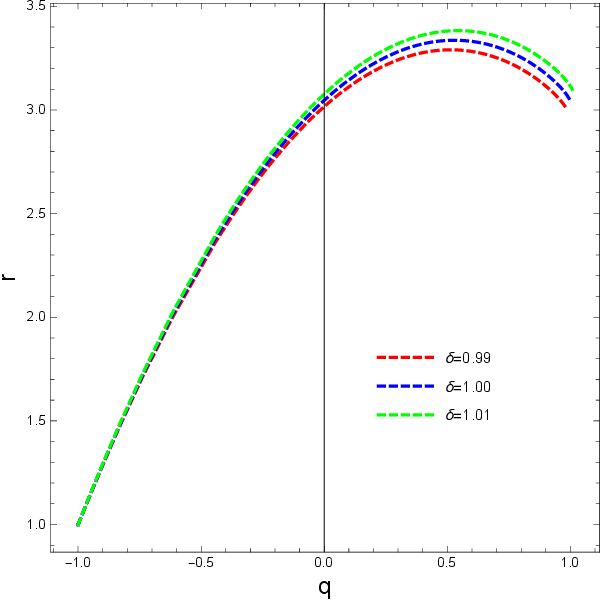}
\caption{Plot of $r$ vs $q$ ( $r-q$ plane) for $f(R,T)=\mu R+\nu T$ model. We have chosen $\rho_{m0} = 0.009$, $\nu = 0.01$, $\mu=0.001$, $\alpha=0.004$, $C_{1}=0.05$, $n=0.02$, $\beta=0.3$.}
\label{F5}
\end{center}
\end{figure}

The evolutionary trajectories of another Statefinder pair $(r,q)$ for interacting THDE with Hubble’s Horizon cutoff in $f(R,T)=\mu R+\nu T$ model is plotted in Fig.\ref{F5}. The fixed points $(q=-1,r=1)$ and point $(q=0.5,r=1)$  in the $q-r$ plane correspond to the de Sitter expansion and the SCDM model. In the $q-r$ plane, $q=1~\&~0$, correspond to matter-dominated era and transition line, respectively, and  $r=1$ represents the evolution of the $\Lambda$CDM model. We can see from the Fig.\ref{F5} that the evolutionary curve of the $q-r$
plane of THDE model starts from the radiation-dominated era $(q=1,r=1)$ in the past and reaches the de Sitter expansion phase $(q=-1,r=1)$ in the future. Thus, the derived models perform like study state models in the future. The recent phase transition of the universe is also revealed by the parameter $q$ as it changes from positive to negative sign.

Another popular diagnostic, $O_{m}(z)$, which is used for comparing various DE models with the $\Lambda$CDM model \cite{41,42}. $O_{m}(z)$ parameter is yielded by the Hubble parameter $H$ and redshift $z$, which is defined by
\begin{equation}
O_{m}(z)=\frac{E(z)-1}{(1+z)^{3}-1},
    \label{E9}
\end{equation}
where $E(z)$ is a dimensionless parameter for rate of expansion which is defined as $E(z)=\frac{H(z)}{H_{0}}$ and $H_{0}$ represents the current value of the Hubble parameter. Based on this tool, DE exhibits the following behaviours: $O_{m}(z)=\Lambda$CDM at its zero curvature, phantom type behaviour at its positive curvature, and quintessence type behaviour at its negative curvature. Despite imprecise knowledge of the matter density, variations in dark energy models can be distinguished by the slope of $O_{m}(z)$. 

\begin{figure}
\begin{center}
\includegraphics[height=3.5in]{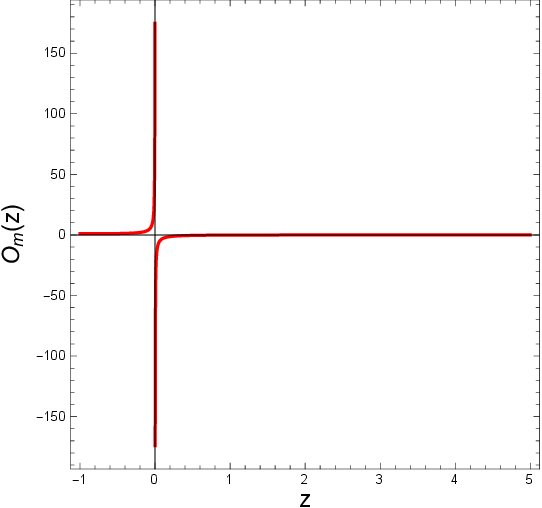}
\caption{Evolution of reconstructed $O_{m}(z)$ diagnostic with respect to redshift $z$ for $f(R,T)=\mu R+\nu T$ model. We have chosen $\rho_{m0} = 0.002$, $\nu = 0.003$, $\mu=0.002$, $\alpha=0.5$, $\delta =0.02$,  $C_{1}=0.4$, $n=0.02$, and $\beta=0.4$.}
\label{F4}
\end{center}
\end{figure}

Deviation from the $\Lambda$CDM is indicated by evolution in the $O_{m}(z)$ diagnostic value. It is easier to be determined
from the present observations, as the $O_{m}(z)$ diagnostic only depends upon the expansion rate. We have the reconstructed $O_{m}(z)$ diagnostic with respect to redshift $z$ for THDE with Hubble's Horizon Cutoff in $f(R,T)$ Gravity as [see Eq.(\ref{E10}) of appendix].
The reconstructed $O_{m}(z)$ diagnostic Eq.(\ref{E10}) for interacting THDE with Hubble’s Horizon cutoff in $f(R,T)=\mu R+\nu T$ model has been plotted in Fig.\ref{F4}. We may be aware that the quintessence behaviour of DE is displayed by the negative curvature of $O_{m}(z)$ trajectories, whereas the positive curvature of $O_{m}(z)$ trajectories resembles phantom behaviour. We can see Fig.\ref{F4} that both regions are provided by the $O_{m}(z)$ trajectories.

\section{Analysis of $w_{DE}-w^{'}_{DE}$ pair}
\begin{figure}
\begin{center}
\includegraphics[height=3.5in]{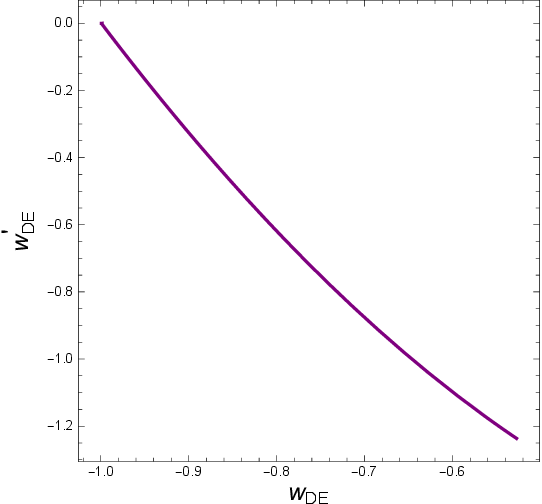}
\caption{Plot of $(w_{DE}-w^{'}_{DE})$ plane for $f(R,T)=\mu R+\nu T$ model. We have chosen $\rho_{m0} = 0.003$, $\nu = 0.009$, $\mu=0.2$, $\alpha=0.035$, $\delta =0.15$,  $C_{1}=0.5$, $n=0.1$, $\beta=0.3$ and $\xi=0.001$.}
\label{F6}
\end{center}
\end{figure}

 Without a doubt, the $w_{DE}-w^{'}_{DE}$ pair evaluation provides us with an optional technique for classifying DE models by using the parameters that represent the dynamical properties of DE. It is evident from this that the Statefinder match $(r,s)$ and the $(w_{DE}-w^{'}_{DE})$ match do indeed identify. Every value of
the Statefinder method states that the parameters of the
Statefinder supports the
scale factor and their differential coefficient, and they are generally pulled out in a model-autonomous way of observational information, despite
the fact that it seems to be problematic to achieve that today, whereas the investigation of the upside of the $w_{DE}-w^{'}_{DE}$ shows
that it holds a dynamical analysis straight towards DE. Consequently, one may consider the dynamical determination of $(w_{DE}-w^{'}_{DE})$ and the geometric conclusion of the Statefinder $(r,s)$ to be rather complementary. In a plane like this, the time or scale factor varies during the phenomenon. The unique structure was identified and the fields were categorized as those that initially slide because of the cosmological constant's steepness (freeze). The limits of the scalar field DE's freezing and thawing models were then examined by \cite{44} in relation to the conflict between the Hubble drag and the potentials' steepness. In Fig.\ref{F6}, we have depicted the $(w_{DE}-w^{'}_{DE})$ plane for interacting THDE with Hubble’s Horizon cutoff in $f(R,T)=\mu R+\nu T$ model. From Fig.\ref{F6}, we can see our model remains in the freezing zone $(w_{DE}<0,w^{'}_{DE}<0)$, which obeys the following observational data \cite{45,46} and the way the trajectories of  $w_{DE}-w^{'}_{DE}$ plane evolve indicates that the cosmic expansion is accelerating more.

\section{$f(R,T)=R+\gamma R^2+\xi T$ Model}
\begin{figure}
\begin{center}
\includegraphics[height=3.5in]{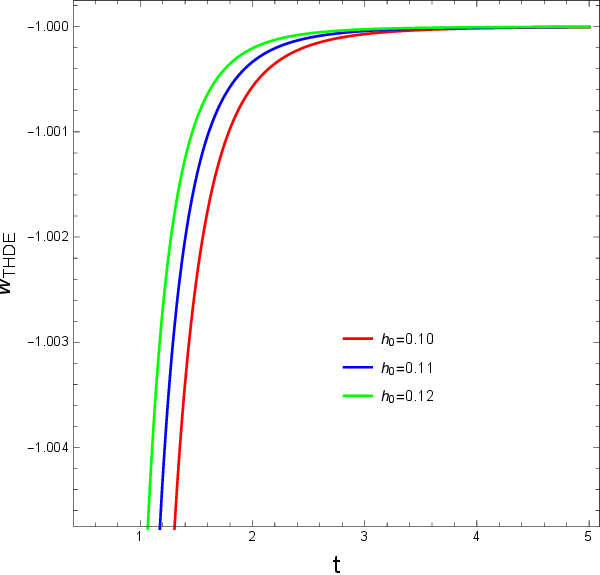}
\caption{Evolution of EoS parameter $w_{THDE}$ with respect to cosmic time $t$ for $f(R,T)=R+\gamma R^2+\xi T$ Model. We have chosen $\rho_{m0} = 0.001$, $\gamma=0.002$, $\Delta = 0.003$, $\alpha=1.1302$, $\delta =0.005$,  $\epsilon_{1}=-1$, $\epsilon_{2}=-1$, $\xi=0.9$, and $\beta=0.004$.}
\label{Q1}
\end{center}
\end{figure}

For comparison with linear form of $f(R,T)$ model, we have considered a polynomial form of $f(R,T)$ model, which is
\begin{equation}
f(R,T)=R+\gamma R^2+\xi T,
    \label{M1}
\end{equation}
where $\gamma$ and $\xi$  are real constants. To derive the equation system of this $f(R,T)$ model, we have chosen a power law form of the scale factor $a$ as
\begin{equation}
    a=\left(1+\frac{h_{0}t}{\beta}(\alpha-1)\right)^{\frac{\beta}{\alpha-1}},
    \label{M2}
\end{equation}
where $h_{0}$, $\alpha$ and $\beta$ are constants and we get the Hubble parameter $H$  as
\begin{equation}
    H=h_{0}\left(1+\frac{h_{0}t}{\beta}(\alpha-1)\right)^{-1},
    \label{M3}
\end{equation}
From Eqs. (\ref{M2}) and (\ref{M3}), we have derived the system of equation for the $f(R,T)=R+\gamma R^2+\xi T$ model and hence the modified field equations for this model have been derived. Through the modified field equations we have reconstructed Hubble parameter $H$ and the THDE density $\rho_{DE}$ Eq.(\ref{E2}) with Hubble horizon for $f(R,T)=R+\gamma R^2+\xi T$ model in the interacting scenario is obtained. By making substantial substitutions in the equation $\dot{H}=-\frac{1}{2}\rho_{DE}(1+w)$, we obtained the EoS parameter $w_{THDE}$ for interacting THDE with Hubble's horizon cutoff in $f(R,T)=R+\gamma R^2+\xi T$ model and it is pictorially represented in Fig. \ref{Q1}.
Fig.\ref{Q1} shows that irrespective of the values of $h_{0}$, EoS parameter is close to $-1$ and it is tending to $-1$ with the evolution of the universe. The behaviour shown is phantom and the EoS parameter never crosses the phantom boundary. We have observed in this polynomial form of $f(R,T)$ model that at $t=5$, $w_{THDE}=-1$ which is consistent with the observations of the current accelerating expansion of the universe. If we compare the EoS parameter presented in Fig.\ref{Q1} with Fig.\ref{F1}, we observe that the EoS parameter is in the neighbourhood of $-1$. However for THDE in $f(R,T)=\mu R+\nu T$, the behaviour of EoS is quintessence and here the crossing of phantom boundary is not realized. Hence we observed that in both cases the transition from quintessence to phantom is not achievable.
\begin{table}[ht!]
\caption{The tabular presentation of the value of EoS parameter for the interacting THDE with Hubble’s Horizon Cutoff in $f(R,T)=R+\gamma R^2+\xi T$ model for the choice of parameters at the point $z=0$ i.e., current accelerating universe.}
\centering
 \begin{tabular}{||c | c | c||} 
 \hline
$(\rho_{m0}, \gamma, \alpha, \Delta, \beta, \xi, \epsilon_{1}, \epsilon_{2}, h_{0}, \delta)$ & $w_{THDE}$ & $w_{Observed} \cite{W0}$\\ [1ex] 
 \hline\hline
 (0.001, 0.002, 1.1302, 0.003, 0.004, 0.9, -1, -1, 0.12, 0.005)  & -1.01 & $-1.06_{-0.13}^{+0.11}$ \\[1ex]
 \hline\hline
 \end{tabular}
 \label{T2}
\end{table}
\begin{figure}
\begin{center}
\includegraphics[height=3.5in]{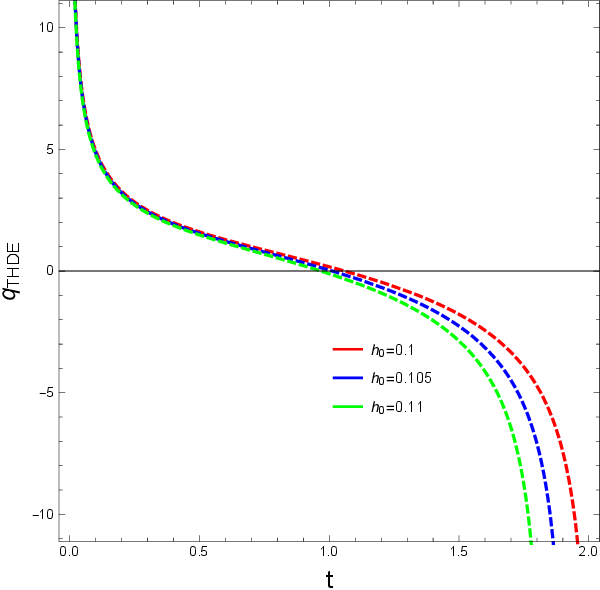}
\caption{Evolution of deceleration parameter $q_{THDE}$ with respect to cosmic time $t$ for $f(R,T)=R+\gamma R^2+\xi T$ model. We have chosen $\rho_{m0} = 0.001$, $\gamma=0.5$, $\Delta = 1.2$, $\alpha=1.1$, $\delta =0.5$, $\epsilon_{1}=-1$, $\epsilon_{2}=1$, $\xi=0.002$, and $\beta=0.9$.}
\label{Q2}
\end{center}
\end{figure}

Let us further comment on deceleration parameter. The evolution of the deceleration parameter $q_{THDE}$ against the redshift $z$ for interacting THDE with Hubble's horizon cutoff in $f(R,T)=R+\gamma R^2+\xi T$ model is pictorially represented in Fiq.\ref{Q2}. Hence we conclude that in case of both the models $f(R,T)=R+\gamma R^2+\xi T$ and $f(R,T)=\mu R+\nu T$, we can obtain accelerated expansion phase of the universe from the decelerated expansion phase and the current accelerated universe is attainable.
\begin{figure}
\begin{center}
\includegraphics[height=3.5in]{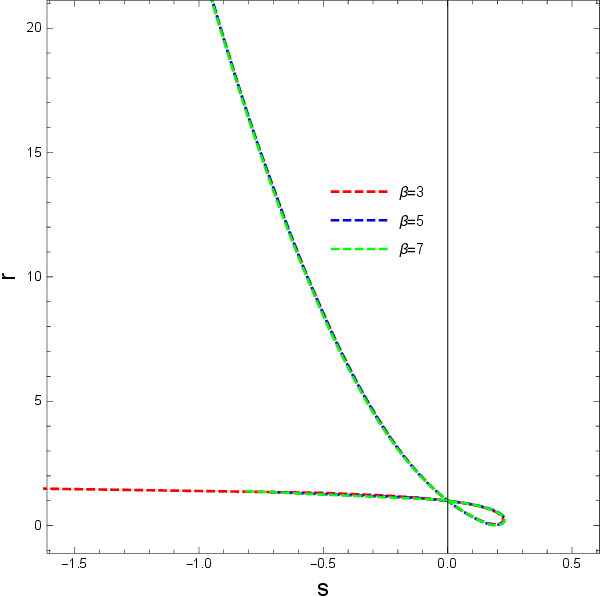}
\caption{The statefinder pair $(r,s)$ trajectory for $f(R,T)=R+\gamma R^2+\xi T$ Model. We have chosen $\rho_{m0} = 0.001$, $\gamma=0.5$, $\Delta = 1$, $\alpha=1.1302$, $\delta =0.005$, $\epsilon_{1}=-1$, $\epsilon_{2}=1$, $h_{0}=0.1$ and $\xi=0.002$.}
\label{Q3}
\end{center}
\end{figure}

In Fig.\ref{Q3}, we have seen that the evolutionary trajectories of Statefinder pair $(r,s)$ for interacting THDE with Hubble's horizon cutoff in $f(R,T)=R+\gamma R^2+\xi T$ model start its evolution from the region $(s<0,r>1)$ and passes through the $\Lambda$CDM fixed point $(s=0,r=1)$ with time. After making a swirl in the quintessence region i.e., $s>0,r<1$, it again passes through the $\Lambda$CDM fixed point, and afterward, it remains in the region $(s<0,r>1)$. As the trajectories pass the $\Lambda$CDM fixed point twice, it affirms that the interacting THDE in $f(R,T)=R+\gamma R^2+\xi T$ model revolves around the $\Lambda$CDM phase of the universe. On the contrary, Fig.\ref{F3} shows that the Statefinder trajectories traverse through the Chaplygin gas phase. Since we are dealing with a late-time universe the $f(R,T)=R+\gamma R^2+\xi T$ model appears to be more realistic than the linear one because of the attainability of the quintessence region. However $\Lambda$CDM phase can be attained in both cases. In the subsequent sections, we will analyze the outcomes through observational data.

\section{Observational Constraints}
This section is a prelude to the observational data analysis in the context of the cosmology presented in the previous sections. Prior to the analysis to be reported subsequent sections, let us have the computational background of the observational data analysis.
From Eq.(\ref{P5}), using Eqs. (\ref{E3}) and (\ref{N1}) we have
\begin{equation}
 \dot{\rho}_{DE}+3H(\rho_{DE}+w_{DE}\rho_{DE})=3H\delta\rho_{m}.
 \label{N3}
\end{equation}
On solving Eq.(\ref{N3}), we have the following expression
\begin{equation}
    \rho_{DE} =\frac{\delta}{w_{DE}-\delta}a^{-3(1+\delta)}+C_{1}a^{-3(1+w_{DE})}.
    \label{N4}
\end{equation}
Now, $\frac{1}{a}=1+z$ which implies $a=(1+z)^{-1}$. Therefore we can rewrite Eqs. (\ref{E3}) and (\ref{N4}) as
\begin{equation}
    \rho_{m}=\rho_{m0}(1+z)^{3(1+\delta)},
    \label{N5}
\end{equation}
\begin{equation}
    \rho_{DE} =\frac{\delta}{w_{DE}-\delta}(1+z)^{3(1+\delta)}+C_{1}(1+z)^{3(1+w_{DE})},
    \label{N6}
\end{equation}
respectively.
From Eqs. (\ref{A13}), (\ref{N5}) and (\ref{N6}) ,we have
\begin{equation}
\begin{array}{cc}
    H^{2}=\frac{\rho_{m0}}{3(\mu+\nu)}(1+z)^{3(1+\delta)}+\frac{\delta}{3(\mu+\nu)(w_{DE}-\delta)}(1+z)^{3(1+\delta)}+\\
    \frac{C_{1}}{3(\mu+\nu)}(1+z)^{3(1+w_{DE})}[\mu\alpha (1+z)^{-n}+\nu\beta (1+z)^{-m}]
\end{array}
\end{equation}
Now we have,
\begin{equation}
    \begin{array}{cc}
    E^{2}(z)=\Omega_{m0}(1+z)^{3(1+\delta)}+\frac{\delta}{3(\mu+\nu)(w_{DE}-\delta)H_{0}^{2}}(1+z)^{3(1+\delta)}+\\
    \Omega_{DE0}(1+z)^{3(1+w_{DE})}[\mu\alpha (1+z)^{-n}+\nu\beta (1+z)^{-m}],
\end{array}
\end{equation}
where $\Omega_{m0}=\frac{\rho_{m0}}{3H_{0}^{2}(\mu+\nu)}$, and $\Omega_{DE0}=\frac{C_{1}}{3H_{0}^{2}(\mu+\nu)}$.

\section{Observational Data Analysis in $f(R,T)$ gravity}

A detailed observational data analysis \cite{B1,B2,B3,B4,B5,B05} using Stern data has been performed in this section for the models considered. Both the models have also been studied under Stern$+$BAO and Stern$+$BAO$+$CMB joint analysis. In the present work the mechanism used is the $\chi^{2}$ minimum test of theoretical Hubble parameter with the observed data set and to find the best fit values of unknown parameters for
different confidence levels (66 $\%$, 90 $\%$, 99 $\%$ ). 

\subsection{Stern $(H(z)-z)$ Data Set}
In this sub-section, we have analyzed the THDE model in $f(R,T)$ gravity using the
observed values of Hubble parameter for different redshifts listed by Stern et al. \cite{B6} in the observed Hubble data set. In Table \ref{T3}, the observed values of Hubble parameter $H(z)$ and the standard error $\sigma(z)$ for different values of redshift $z$ are given. A
statistical hypothesis is put forth and at different confidence levels its validity is tested.
\begin{table}[ht!]
\caption{The tabular presentation of the Hubble parameter
$H(z)$ and the standard error $\sigma(z)$ for different values of redshift $z$.}
\centering
 \begin{tabular}{||c | c | c||} 
 \hline
~~~~~$z$~~~~~ & ~~~~~$H(z)$~~~~~ & ~~~~~$\sigma(z)$~~~~~\\ [1ex] 
 \hline\hline
 0 & 73 & $\pm8$ \\[1ex]
 \hline
 0.1 & 69 & $\pm12$ \\[1ex]
 \hline
 0.17 & 83 & $\pm8$ \\[1ex]
 \hline
 0.27 & 77 & $\pm14$ \\[1ex]
 \hline
 0.4 & 95 & $\pm17.4$ \\[1ex]
 \hline
 0.48 & 90 & $\pm60$ \\[1ex]
 \hline
 0.88 & 97 & $\pm40.4$ \\[1ex]
 \hline
 0.9 & 117 & $\pm23$ \\[1ex]
 \hline
 1.3 & 168 & $\pm17.4$ \\[1ex]
 \hline
 1.43 & 177 & $\pm18.2$ \\[1ex]
 \hline
 1.53 & 140 & $\pm14$ \\[1ex]
 \hline
 1.75 & 202 & $\pm40.4$ \\[1ex]
 \hline\hline
 \end{tabular}
 \label{T3}
\end{table}
To achieve this, we first create the $\chi^{2}$ statistics as a sum of standard normal distribution as follows:
\begin{equation}
\chi^{2}_{Stern}=\sum\frac{(H(z)-H_{obs}(z))^{2}}{\sigma^
{2}(z)}
\label{M03}
\end{equation}
\begin{equation}
L=\int e^{-\frac{1}{2}\chi^{2}_{Stern}}P(H_{0})dH_{0},
    \label{M4}
\end{equation}
where $H(z)$ and $H_{obs}(z)$ are theoretical and observational values of the Hubble parameter for different values of redshifts respectively. $H_{obs}(z)$ can be safely marginalized in this case as it is a nuisance parameter. The present value of Hubble parameter is fixed at $H_{0} = 72 \pm 8$ Km$s^{-1}$ Mp$c^{-1}$. We can determine best fit value of the parameters ($\beta$ vs  $\alpha$)  by minimizing the abovementioned distribution $\chi^{2}_{Stern}$ and fixing the remaining unknown parameters with the help of Stern data. 
In Table \ref{T4} , the best fit values of $\beta$ against  $\alpha$ are presented according to our analysis. For different confidence levels, the graph is plotted. The theoretical range of the parameters is supported by our best fit analysis using Stern observational data.
\begin{table}[ht!]
\caption{The best fit values of $\alpha$,
$\beta$ and the minimum values of $\chi^{2}$}
\centering
 \begin{tabular}{||c | c | c | c ||} 
 \hline
~~~~~~Data~~~~~~ & ~~~~~~$\alpha$~~~~~~ & ~~~~~~$\beta$~~~~~~&~~~~~~$\chi^{2}_{min}$~~~~~~\\ [1ex] 
 \hline\hline
 Stern & 616.715 & -10 & 131.313 \\[1ex]
 \hline
 Stern+BAO & 610.43 & -1 & 892.088 \\[1ex]
 \hline
 Stern+BAO+CMB & 614.93 & -9.99995 & 10086.8 \\[1ex]
 \hline\hline
 \end{tabular}
 \label{T4}
\end{table}
\begin{figure}
\begin{center}
\includegraphics[height=3.5in]{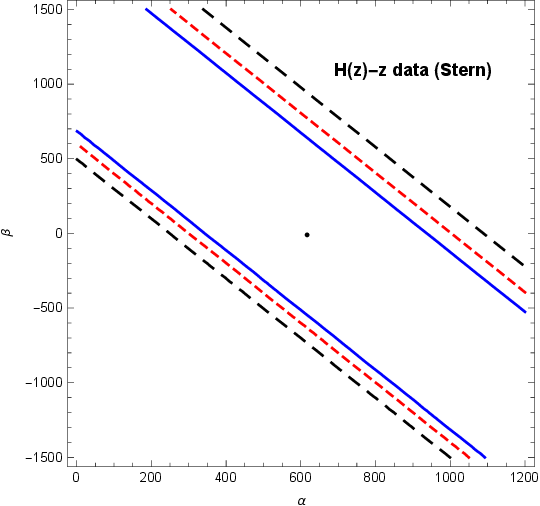}
\caption{The variations of $\beta$ against $\alpha$ for the $(H(z)-z)$ joint analysis. The parameters chosen are $\Omega_{m0}=0.01$, $\Omega_{DE0}=0.02$, $\delta=1.5$, $w_{DE}=0.01$, $n=0.1$, $m=0.1$, $\mu=0.02$, $\nu=0.01$, $H_{0}=72$. The contours are drawn for 66 $\%$
(solid, blue), 90 $\%$ (dashed, red) and 99 $\%$ (dashed, black) confidence levels.}
\label{Q4}
\end{center}
\end{figure}
In Fig. \ref{Q4}, we have plotted the contours of $\beta$ vs $\alpha$ for 66 $\%$
(solid, blue), 90 $\%$ (dashed, red) and 99 $\%$ (dashed, black) confidence levels.

\subsection{Stern $+$ BAO Data Sets}
In this subsection, a joint analysis is resorted in the sense that the stern data incorporates BAO peaks. \cite{B7} proposed the Baryon Acoustic Oscillation (BAO) peak parameter value and their approach has been used over here. The Sloan Digital Sky Survey (SDSS) survey is regarded as the pioneer as far as the BAO signal detection is concerned. At a scale $\sim$ 100 MPc, the survey directly detects BAO signals. The analysis combines angular diameter distance and Hubble parameter at a particular redshift. This analysis contains no specific dark energy and is independent of the $H_{0}$ measurement. Here, we use standard $chi^{2}$ analysis to investigate the parameters $\beta$ vs $\alpha$ for the interacting THDE model from the data of the BAO peak for low redshift (with range $0<z<0.35$). When a Gaussian distribution is considered, the error is equivalent to the standard deviation. We are aware that the Low-redshift distance measurements can measure the Hubble constant $H_{0}$ directly and are very lightly dependent on various cosmological constants and the EoS of dark energy. The BAO peak parameter is defined as 
\begin{equation}
    \mathcal{A}=\frac{\sqrt{\Omega_{m}}}{E(z_{1})^{\frac{1}{3}}}\biggl(\frac{1}{z_{1}}\int_{0}^{z_{1}}\frac{dz}{E(z)}\biggr)^{\frac{2}{3}}.
    \label{M5}
\end{equation}
Here $E(z)$ is the normalized Hubble parameter, the integration term is the dimensionless comoving distance for the redshift $z_{1}$ and the redshift $z_{1}=0.35$ is the
typical redshift of the SDSS sample. For the flat model of the universe, the value of the parameter $\mathcal{A}$ is given by $\mathcal{A}=0.469 \pm 0.017$ using SDSS data \cite{B7} from the survey of luminous red galaxies . The $\chi^{2}$ function for the BAO measurement is 
\begin{equation}
    \chi^{2}_{BAO}=\frac{(\mathcal{A}-0.469)^{2}}{(0.017)^{2}}.
    \label{M6}
\end{equation}
The total joint analysis of (Stern $+$ BAO) data sets for the $\chi^{2}$ function is defined by
\begin{equation}
    \chi^{2}_{Total}= \chi^{2}_{Stern}+ \chi^{2}_{BAO}.
    \label{M7}
\end{equation}
\begin{figure}
\begin{center}
\includegraphics[height=3.5in]{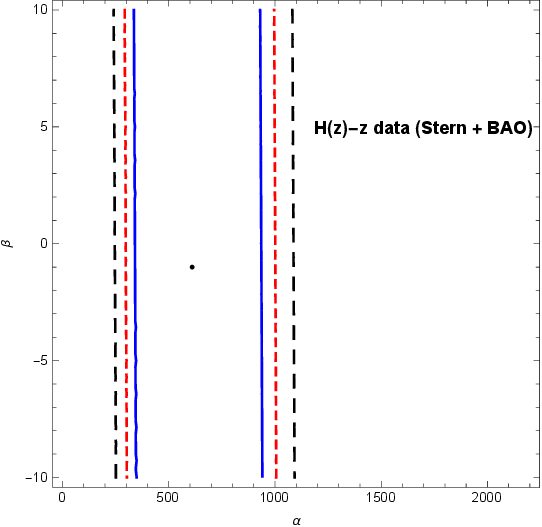}
\caption{The variations of $\beta$ against $\alpha$ for the $(H(z)-z)+ BAO$ joint analysis. The parameters chosen are $\Omega_{m0}=0.01$, $\Omega_{DE0}=0.02$, $\delta=1.5$, $w_{DE}=0.01$, $n=0.1$, $m=0.1$, $\mu=0.02$, $\nu=0.01$, $H_{0}=72$. The contours are drawn for 66 $\%$
(solid, blue), 90 $\%$ (dashed, red) and 99 $\%$ (dashed, black) confidence levels.}
\label{Q5}
\end{center}
\end{figure}
As per our analysis, the best fit values of $\beta$ vs $\alpha$ for the joint data analysis of $(H(z)-z+ BAO)$ is presented
in Table \ref{T4}. In Fig. \ref{Q5}, we have plotted the closed contours of $\beta$ vs $\alpha$  for 66 $\%$
(solid, blue), 90 $\%$ (dashed, red) and 99 $\%$ (dashed, black) confidence levels.

\subsection{Stern $+$ BAO $+$ CMB Data Sets}
The angular scale of the sound horizon at the last scattering surface is used to calculate the angular scale of the first acoustic peak. This is one of the most intriguing geometrical investigations of dark energy. The CMB (Cosmic Microwave Background) power spectrum contains the information. \cite{B8,B9,B10} defines the CMB shift parameter. It can be used to confine the model parameters, although it is not responsive to perturbations. This property has been used in this analysis. The CMB power spectrum's first peak is the shift parameter, which is given by
\begin{equation}
\mathcal{R}=\sqrt{\Omega_{m}}\int^{z_{2}}_{0}\frac{dz}{E(z)},
    \label{M8}
\end{equation}
where $z_{2}$ is the value of redshift at the surface of the last scattering. Based on Komatsu et al.'s work \cite{B11} using WMAP7 data, the parameter's value was determined to be  $R=1.726 \pm 0.018$ at redshift $z=1091.3$. The CMB measurement's $\chi^{2}$ function can be expressed as
    \begin{equation}
    \chi^{2}_{CMB}=\frac{(\mathcal{R}-1.726)^{2}}{(0.018)^{2}}.
    \label{M9}
\end{equation}
The total joint analysis of (Stern $+$ BAO $+$ CMB) data sets  for the $\chi^{2}$ function is defined as
\begin{equation}
    \chi^{2}_{Total}= \chi^{2}_{Stern} + \chi^{2}_{BAO} + \chi^{2}_{CMB} .
    \label{M10}
\end{equation}
\begin{figure}
\begin{center}
\includegraphics[height=3.5in]{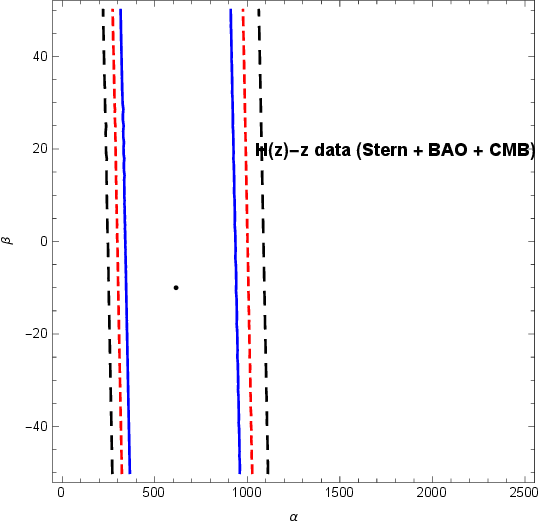}
\caption{The variations of $\beta$ against $\alpha$ for the $(H(z)-z)+ BAO + CMB$ joint analysis. The parameters chosen are $\Omega_{m0}=0.01$, $\Omega_{DE0}=0.02$, $\delta=1.5$, $w_{DE}=0.01$, $n=0.1$, $m=0.1$, $\mu=0.02$, $\nu=0.01$, $H_{0}=72$. The contours are drawn for 66 $\%$
(solid, blue), 90 $\%$ (dashed, red) and 99 $\%$ (dashed, black) confidence levels.}
\label{Q6}
\end{center}
\end{figure}
As per our analysis, the best-fit values of $\beta$ vs. $\alpha$ for the joint data analysis of $(Stern $+$ BAO $+$ CMB)$ which support the theoretical range of the parameters is presented in Table \ref{T4}. In Fig. \ref{Q6}, we have plotted the closed contours of $\beta$ vs $\alpha$  for 66 $\%$
(solid, blue), 90 $\%$ (dashed, red) and 99 $\%$ (dashed, black) confidence levels.

\subsection{Supernovae Type Ia: Redshift-Magnitude Observations}
Supernova Type Ia experiments provide the primary evidence for dark energy's existence. The universe's redshift directly relates to the existence of dark energy. Two Supernova Cosmology Project and High Redshift Supernova Search teams have worked hard since 1995. As a result of their efforts, they have found multiple type Ia supernovae at high redshifts \cite{C5,B11,B12,B13}. The distance modulus of a Supernovae and its redshift $z$ \cite{B12,B14} is measured directly by the observations. In this case, we will consider the most current observational data, such as SNe Ia, which is part of the Union2 sample \cite{B15} and has 557 data points. From the observations, the dark energy density determines the luminosity distance $d_{L}(z)$ and is defined as
\begin{equation}
    d_{L}(z)=(1+z)H_{0}\int^{z}_{0} \frac{dz^{'}}{H(z^{'})}.
    \label{M11}
\end{equation}
The distance modulus for Supernovae is defined as the distance between absolute and apparent luminosity of a distance
object and is given by
\begin{equation}
    \mu(z)=5 log_{10}\biggl[\frac{d_{L}(z)/ H_{0}}{1 MPc}\biggr]+25.
    \label{M12}
\end{equation}
\begin{figure}
\begin{center}
\includegraphics[height=3.5in]{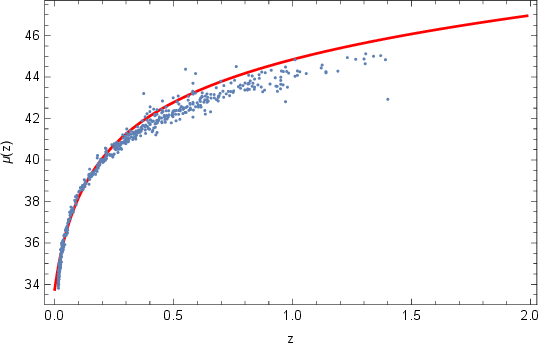}
\caption{Plot of $\mu(z)$ vs $z$. The parameters chosen are $\Omega_{m0}=0.01$, $\Omega_{DE0}=0.02$, $\delta=1.5$, $w_{DE}=0.01$, $n=0.1$, $m=0.1$, $\mu=0.02$, $\nu=0.01$, $H_{0}=72$.}
\label{Q7}
\end{center}
\end{figure}
In Fig.\ref{Q7}, the best fit of distance modulus $\mu(z)$ as a function  of redshift $z$ for our model and the Supernova Type Ia Union2 sample has been plotted for the best fit values of the parameters considered. From the plot, we can conclude that the interacting THDE model in $f(R,T)$ gravity is in good agreement with the union2 sample data.

\section{Concluding Remarks}
In this work, we have studied the cosmological consequences of interacting THDE model, a particular case of a more generalized Nojiri-Odintsov HDE \cite{Nojiri:2005pu, Nojiri:2017opc} with Hubble’s horizon cutoff in $f(R,T)=\mu R+\nu T$ model. We have observed that the EoS parameter $w_{DE}$ exhibits quintessence behaviour and at $z=0$, $w_{DE}\approx -0.99$ which is in good accordance with the current cosmological observations \cite{W0} [see Table \ref{T1}]. A similar behaviour is seen in the total EoS parameter $w_{DE,total}$. The deceleration parameter $q$ shows a transition from positive to negative sign, i.e., the universe is gradually evolving from the decelerated expansion phase $(q > 0)$ to the accelerated expansion phase $(q < 0)$. The evolutionary trajectories of the reconstructed Statefinder pair $(r,s)$  pass the $\Lambda$CDM fixed point twice, which confirms strongly that the interacting THDE in the model $f(R,T)=\mu R+\nu T$ circulates the $\Lambda$CDM phase of the universe. The  $q-r$ trajectories of interacting THDE in $f(R,T)=\mu R+\nu T$ model evolves from the radiation-dominated era $(q=1,r=1)$ in the past to the de Sitter expansion phase $(q=-1,r=1)$ in the future. As the parameter $q$ changes from a positive to a negative sign, it depicts the recent phase transition of the universe. The reconstructed $O_{m}(z)$ diagnostic exhibits both the quintessence behaviour of DE which is displayed by the negative curvature of $O_{m}(z)$ trajectories and the phantom behaviour of DE which is displayed by the positive curvature of $O_{m}(z)$ trajectories. In $(w_{DE}-w^{'}_{DE})$ pair analysis, we have seen our model $f(R,T)=\mu R+\nu T$ remains in the freezing zone $(w_{DE}<0,w^{'}_{DE}<0)$, which obeys the following observational data \cite{45,46} and the evolution of the trajectory of  $w_{DE}-w^{'}_{DE}$ plane indicates more acceleration in the cosmic expansion.

Interacting THDE model with Hubble’s horizon cutoff has also been studied in $f(R,T)=R+\gamma R^2+\xi T$ model. In case of this polynomial form of $f(R,T)$ model, we have seen that the EoS parameter $w_{THDE}$ shows phantom behaviour and it is tending to $-1$ with the evolution of the universe but never crosses the phantom boundary and at $z=0$, $w_{THDE}\approx -1.01$ which is the best fit value \cite{W0} [see Table \ref{T2}]. In case of the deceleration parameter $q_{THDE}$, the accelerated expansion phase of the universe from the decelerated expansion phase has been achieved. In this scenario, the evolutionary trajectories of Statefinder pair $(r,s)$ revolves around the $\Lambda$CDM phase of the universe.

While comparing the linear and polynomial forms of the $f(R,T)$ model, we have encountered that the EoS parameter in case of $f(R,T)=\mu R+\nu T$ model exhibits quintessence behaviour. On the contrary, the EoS parameter in case of $f(R,T)=R+\gamma R^2+\xi T$ model shows phantom behaviour. We have seen that the current accelerated universe is attainable in both models considered. The Statefinder pair $(r,s)$ trajectories for $f(R,T)=R+\gamma R^2+\xi T$ model appears to be more realistic than the $f(R,T)=\mu R+\nu T$ model because of the attainability of the quintessence region but $\Lambda$CDM fixed point has been achieved for both the models under consideration.

We have developed observational constraints and analyzed the interacting THDE model in $f(R,T)$ gravity with Stern $(H(z)-z)$ data set, Stern $+$ BAO data sets and Stern $+$ BAO $+$ CMB data sets. The best-fit values of model parameters $\alpha$ and $\beta$ have been obtained through $\chi^{2}$ minimum technique, and the results have been presented in Table \ref{T4}. We have plotted the statistical confidence contours of $\beta$ vs $\alpha$  for 66 $\%$
(solid, blue), 90 $\%$ (dashed, red) and 99 $\%$ (dashed, black) confidence levels by fixing remaining observable parameters for Stern data analysis in Fig.\ref{Q4}, Stern$+$BAO joint data analysis in Fig.\ref{Q5} and Stern$+$BAO$+$CMB joint data analysis in Fig.\ref{Q6}. We compare the model parameters for each scenario using both the statistical contours and their values. The comparative analysis provides insight into how the theoretical parameter values converge with the values derived from the observational data set and how these values change for various selected group of additional parametric values. Finally, we have plotted the distance modulus $\mu(z)$ against redshift $z$ in Fig.\ref{Q7} for the interacting THDE model in $f(R,T)$ gravity for the best fit values of the observable parameters 
and the observed Supernova Type Ia Union2 data sample. Hence, the observational data sets are in good accordance with our predicted theoretical interacting THDE model in $f(R,T)$ gravity.

We believe it is premature to constrain ourselves to small values of the parameter $\delta$ because the origin and behaviour of DE are not well understood and also due to GR's shortcomings in characterizing DE, and the inability of HDE-based Bekenstein entropy to explain DE. To determine and utilise this constraint, further observations and research \cite{48,49,C68,50} are still required. In conclusion, the Tsallis holographic dark energy scenario, as measured by the new parameter $\delta$, shows richer behaviour than the standard holographic dark energy scenario. However, because of its consistent formulation, the standard holographic dark energy scenario can still be obtained as a sub-case, specifically for $\delta=1$. Before the scenario is taken seriously as a viable contender for the description of nature, more research \cite{51} has to be done. To summarise, the discussion concludes with the suggestion that, despite the fact that the quantum aspect of gravity has little impact on the observational constraint, cosmological observations can place upper bounds on the magnitude of the correction resulting from quantum gravity, which may be closer to theory than one would anticipate. We would also like to address tension problems for the interacting models considered in our future study.

\section{Acknowledgement}
The authors express sincere gratitude to the anonymous reviewers for their constructive comments. The authors acknowledge the hospitality of the Inter-University Centre for Astronomy and Astrophysics (IUCAA), Pune, India, during their scientific visit in December 2023-January 2024.

\section{Appendix} 
\begin{equation}
\begin{array}{c}
H=\biggl( 3 (\mu +\nu )^2-\biggl(0.104993 a^{-6-6\delta } (\mu +\nu
) \left(288 a^{12 +12\delta } C_{1}+288 a^{9+9\delta } \rho_{m0} 
-1296 a^{12+12\delta } (\mu +\nu )^3-\right.\\
\left. 144 a^{12+n+12 \delta } (\alpha  \mu +\beta  \nu )\biggr) \biggl( 1728 a^{18+18\delta} C_{1}^{2}+3456
a^{15 +15 \delta} C_{1} \rho_{m0}+1728 a^{12+12\delta } \rho_{m0}^2 \right.\\
 -31104 a^{18+18\delta } C_{1} (\mu +\nu )^{3}-31104 a^{15+15\delta } \rho_{m0} (\mu +\nu )^{3}+\\
93312  a^{18 +18\delta } (\mu +\nu )^{6}-1728 a^{18 +n+18\delta } C_{1} (\alpha  \mu +\beta  \nu )-\\
1728 a^{15 +n+15\delta } \rho_{m0} (\alpha  \mu +\beta  \nu )+15552 a^{18+n+18 \delta } (\mu +\nu )^{3} (\alpha  \mu +\beta  \nu )+\\ 432 a^{18+2n+18 \delta } (\alpha  \mu +\beta  \nu )^{2}+ \left(4
(\mu +\nu )^{3} \left(288 a^{12+12\delta } C_{1}+288 a^{9+9 \delta } \rho_{m0}-\right.\right.\\
\left. 1296 a^{12+12 \delta } (\mu +\nu )^{3}-144 a^{12 +n+12 \delta } (\alpha  \mu +\beta  \nu )\right)^{3}+\left(1728
a^{18+18 \delta } C_{1}^2+3456 a^{15+15\delta } C_{1}
\rho_{m0}+\right.\\
1728 a^{12 +12\delta } \rho_{m0}^{2}-
31104 a^{18 +18 \delta } C_{1} (\mu +\nu )^{3}-31104 a^{15+15\delta } \rho_{m0} (\mu +\nu )^{3}+\\
93312 a^{18 +18 \delta } (\mu +\nu )^{6}-1728 a^{18 +n+18
\delta } C_{1} (\alpha  \mu +\beta  \nu )-\\
1728  a^{15 +n+15\delta } \rho_{m0} (\alpha  \mu +\beta  \nu )+15552 a^{18+n+18 \delta } (\mu +\nu )^{3} (\alpha  \mu +\beta  \nu )+\\
\left.\left. 432 a^{18+2 n+18 \delta } (\alpha  \mu +\beta  \nu )^{2} \right)^{2}\right)^{\frac{1}{2}}\biggr)^{-1/3}+0.0661417
a^{-6-6 \delta } \left(1728 a^{18 +18 \delta } C_{1}^{2}+3456
a^{15+15 \delta } C_{1} \rho_{m0}+\right.\\
1728 a^{12 +12\delta} \rho_{m0}^2-
31104 a^{18 +18 \delta } C_{1} (\mu +\nu )^{3}-31104 a^{15+15\delta } \rho_{m0} (\mu +\nu )^{3}+93312 a^{18 +18\delta } (\mu +\nu )^{6}-\\
1728 a^{18 +n+18 \delta } C_{1} (\alpha  \mu +\beta  \nu )-1728 a^{15
+n+15\delta } \rho_{m0} (\alpha  \mu +\beta  \nu )+\\
15552 a^{18+n+18\delta } (\mu +\nu )^{3} (\alpha  \mu +\beta  \nu )+432 a^{18
+2 n+18 \delta } (\alpha  \mu +\beta  \nu )^{2}+ \left(4 (\mu +\nu )^{3} \left(288
a^{12 +12\delta } C_{1}+\right.\right.\\
288 a^{9 +9\delta } \rho_{m0}-1296
a^{12 +12 \delta } (\mu +\nu )^{3}-\\
\left.144 a^{12 +n+12\delta } (\alpha  \mu +\beta  \nu )\right)^{3}+\left(1728 a^{18 +18 \delta } C_{1}^2+3456 a^{15 +15 \delta } C_{1}
\rho_{m0}+1728 a^{12 +12 \delta } \rho_{m0}^{2}-\right.\\
31104 a^{18 +18 \delta } C_{1} (\mu +\nu )^3-31104 a^{15
+15\delta } \rho_{m0} (\mu +\nu )^3+93312 a^{18 +18\delta } (\mu +\nu )^{6} -\\
1728 a^{18 +n+18 \delta } C_{1} (\alpha  \mu +\beta  \nu )-1728 a^{15
+n+15 \delta } \rho_{m0} (\alpha  \mu +\beta  \nu )+\\
\left.\left. 15552 a^{18 +n+18 \delta } (\mu +\nu )^3 (\alpha  \mu +\beta  \nu
)+432 a^{18 +2 n+18\delta } (\alpha  \mu +\beta  \nu )^2\right)^2 \biggr)^{\frac{1}{2}}\right)^{1/3}\biggr)^{\frac{1}{2}}.
\end{array}
    \label{E4}
\end{equation}
\begin{equation}
    \begin{array}{c}
\rho_{DE}=\biggl( 3 (\mu +\nu )^2-\biggl(0.104993 a^{-6-6\delta } (\mu +\nu
) \left(288 a^{12 +12\delta } C_{1}+288 a^{9+9\delta } \rho_{m0} 
-1296 a^{12+12\delta } (\mu +\nu )^3-\right.\\
\left. 144 a^{12+n+12 \delta } (\alpha  \mu +\beta  \nu )\biggr) \biggl( 1728 a^{18+18\delta} C_{1}^{2}+3456
a^{15 +15 \delta} C_{1} \rho_{m0}+1728 a^{12+12\delta } \rho_{m0}^2 \right.\\
 -31104 a^{18+18\delta } C_{1} (\mu +\nu )^{3}-31104 a^{15+15\delta } \rho_{m0} (\mu +\nu )^{3}+\\
93312  a^{18 +18\delta } (\mu +\nu )^{6}-1728 a^{18 +n+18\delta } C_{1} (\alpha  \mu +\beta  \nu )-\\
1728 a^{15 +n+15\delta } \rho_{m0} (\alpha  \mu +\beta  \nu )+15552 a^{18+n+18 \delta } (\mu +\nu )^{3} (\alpha  \mu +\beta  \nu )+\\ 432 a^{18+2n+18 \delta } (\alpha  \mu +\beta  \nu )^{2}+ \left(4
(\mu +\nu )^{3} \left(288 a^{12+12\delta } C_{1}+288 a^{9+9 \delta } \rho_{m0}-\right.\right.\\
\left. 1296 a^{12+12 \delta } (\mu +\nu )^{3}-144 a^{12 +n+12 \delta } (\alpha  \mu +\beta  \nu )\right)^{3}+\left(1728
a^{18+18 \delta } C_{1}^2+3456 a^{15+15\delta } C_{1}
\rho_{m0}+\right.\\
1728 a^{12 +12\delta } \rho_{m0}^{2}-
31104 a^{18 +18 \delta } C_{1} (\mu +\nu )^{3}-31104 a^{15+15\delta } \rho_{m0} (\mu +\nu )^{3}+\\
93312 a^{18 +18 \delta } (\mu +\nu )^{6}-1728 a^{18 +n+18
\delta } C_{1} (\alpha  \mu +\beta  \nu )-\\
1728  a^{15 +n+15\delta } \rho_{m0} (\alpha  \mu +\beta  \nu )+15552 a^{18+n+18 \delta } (\mu +\nu )^{3} (\alpha  \mu +\beta  \nu )+\\
\left.\left. 432 a^{18+2 n+18 \delta } (\alpha  \mu +\beta  \nu )^{2} \right)^{2}\right)^{\frac{1}{2}}\biggr)^{-1/3}+0.0661417
a^{-6-6 \delta } \left(1728 a^{18 +18 \delta } C_{1}^{2}+3456
a^{15+15 \delta } C_{1} \rho_{m0}+\right.\\
1728 a^{12 +12\delta} \rho_{m0}^2-
31104 a^{18 +18 \delta } C_{1} (\mu +\nu )^{3}-31104 a^{15+15\delta } \rho_{m0} (\mu +\nu )^{3}+93312 a^{18 +18\delta } (\mu +\nu )^{6}-\\
1728 a^{18 +n+18 \delta } C_{1} (\alpha  \mu +\beta  \nu )-1728 a^{15
+n+15\delta } \rho_{m0} (\alpha  \mu +\beta  \nu )+\\
15552 a^{18+n+18\delta } (\mu +\nu )^{3} (\alpha  \mu +\beta  \nu )+432 a^{18
+2 n+18 \delta } (\alpha  \mu +\beta  \nu )^{2}+ \left(4 (\mu +\nu )^{3} \left(288
a^{12 +12\delta } C_{1}+\right.\right.\\
288 a^{9 +9\delta } \rho_{m0}-1296
a^{12 +12 \delta } (\mu +\nu )^{3}-\\
\left.144 a^{12 +n+12\delta } (\alpha  \mu +\beta  \nu )\right)^{3}+\left(1728 a^{18 +18 \delta } C_{1}^2+3456 a^{15 +15 \delta } C_{1}
\rho_{m0}+1728 a^{12 +12 \delta } \rho_{m0}^{2}-\right.\\
31104 a^{18 +18 \delta } C_{1} (\mu +\nu )^3-31104 a^{15
+15\delta } \rho_{m0} (\mu +\nu )^3+93312 a^{18 +18\delta } (\mu +\nu )^{6} -\\
1728 a^{18 +n+18 \delta } C_{1} (\alpha  \mu +\beta  \nu )-1728 a^{15
+n+15 \delta } \rho_{m0} (\alpha  \mu +\beta  \nu )+\\
\left.\left. 15552 a^{18 +n+18 \delta } (\mu +\nu )^3 (\alpha  \mu +\beta  \nu
)+432 a^{18 +2 n+18\delta } (\alpha  \mu +\beta  \nu )^2\right)^2 \biggr)^{\frac{1}{2}}\right)^{1/3}\biggr)^{2-\xi}.
    \label{E5}
    \end{array}
\end{equation}
\begin{equation}
\begin{array}{c}
O_{m}(z)=\frac{1}{-1+(1+z)^3}\Biggl(-1+\frac{1}{H_{0}^2}\biggl( 3 (\mu +\nu )^2-\biggl(0.104993 a^{-6-6\delta } (\mu +\nu
) \left(288 a^{12 +12\delta } C_{1}+288 a^{9+9\delta } \rho_{m0} -\right.\\
1296 a^{12+12\delta }\left.(\mu +\nu )^3- 144 a^{12+n+12 \delta } (\alpha  \mu +\beta  \nu )\biggr) \biggl( 1728 a^{18+18\delta} C_{1}^{2}+3456
a^{15 +15 \delta} C_{1} \rho_{m0}+ \right.\\
1728 a^{12+12\delta } \rho_{m0}^2 -31104 a^{18+18\delta } C_{1} (\mu +\nu )^{3}-31104 a^{15+15\delta } \rho_{m0} (\mu +\nu )^{3}+\\
93312  a^{18 +18\delta } (\mu +\nu )^{6}-1728 a^{18 +n+18\delta } C_{1} (\alpha  \mu +\beta  \nu )-\\
1728 a^{15 +n+15\delta } \rho_{m0} (\alpha  \mu +\beta  \nu )+15552 a^{18+n+18 \delta } (\mu +\nu )^{3} (\alpha  \mu +\beta  \nu )+\\ 432 a^{18+2n+18 \delta } (\alpha  \mu +\beta  \nu )^{2}+ \left(4
(\mu +\nu )^{3} \left(288 a^{12+12\delta } C_{1}+288 a^{9+9 \delta } \rho_{m0}-\right.\right.\\
\left. 1296 a^{12+12 \delta } (\mu +\nu )^{3}-144 a^{12 +n+12 \delta } (\alpha  \mu +\beta  \nu )\right)^{3}+\left(1728
a^{18+18 \delta } C_{1}^2+3456 a^{15+15\delta } C_{1}
\rho_{m0}+\right.\\
1728 a^{12 +12\delta } \rho_{m0}^{2}-
31104 a^{18 +18 \delta } C_{1} (\mu +\nu )^{3}-31104 a^{15+15\delta } \rho_{m0} (\mu +\nu )^{3}+\\
93312 a^{18 +18 \delta } (\mu +\nu )^{6}-1728 a^{18 +n+18
\delta } C_{1} (\alpha  \mu +\beta  \nu )-\\
1728  a^{15 +n+15\delta } \rho_{m0} (\alpha  \mu +\beta  \nu )+15552 a^{18+n+18 \delta } (\mu +\nu )^{3} (\alpha  \mu +\beta  \nu )+\\
\left.\left. 432 a^{18+2 n+18 \delta } (\alpha  \mu +\beta  \nu )^{2} \right)^{2}\right)^{\frac{1}{2}}\biggr)^{-1/3}+0.0661417
a^{-6-6 \delta } \left(1728 a^{18 +18 \delta } C_{1}^{2}+3456
a^{15+15 \delta } C_{1} \rho_{m0}+\right.\\
1728 a^{12 +12\delta} \rho_{m0}^2-
31104 a^{18 +18 \delta } C_{1} (\mu +\nu )^{3}-31104 a^{15+15\delta } \rho_{m0} (\mu +\nu )^{3}+93312 a^{18 +18\delta } (\mu +\nu )^{6}-\\
1728 a^{18 +n+18 \delta } C_{1} (\alpha  \mu +\beta  \nu )-1728 a^{15
+n+15\delta } \rho_{m0} (\alpha  \mu +\beta  \nu )+\\
15552 a^{18+n+18\delta } (\mu +\nu )^{3} (\alpha  \mu +\beta  \nu )+432 a^{18
+2 n+18 \delta } (\alpha  \mu +\beta  \nu )^{2}+ \left(4 (\mu +\nu )^{3} \left(288
a^{12 +12\delta } C_{1}+\right.\right.\\
288 a^{9 +9\delta } \rho_{m0}-1296
a^{12 +12 \delta } (\mu +\nu )^{3}-\\
\left.144 a^{12 +n+12\delta } (\alpha  \mu +\beta  \nu )\right)^{3}+\left(1728 a^{18 +18 \delta } C_{1}^2+3456 a^{15 +15 \delta } C_{1}
\rho_{m0}+1728 a^{12 +12 \delta } \rho_{m0}^{2}-\right.\\
31104 a^{18 +18 \delta } C_{1} (\mu +\nu )^3-31104 a^{15
+15\delta } \rho_{m0} (\mu +\nu )^3+93312 a^{18 +18\delta } (\mu +\nu )^{6} -\\
1728 a^{18 +n+18 \delta } C_{1} (\alpha  \mu +\beta  \nu )-1728 a^{15
+n+15 \delta } \rho_{m0} (\alpha  \mu +\beta  \nu )+\\
\left.\left. 15552 a^{18 +n+18 \delta } (\mu +\nu )^3 (\alpha  \mu +\beta  \nu
)+432 a^{18 +2 n+18\delta } (\alpha  \mu +\beta  \nu )^2\right)^2 \biggr)^{\frac{1}{2}}\right)^{1/3}\Biggr)\Biggr).
\end{array}
\label{E10}
\end{equation}
\end{document}